\documentclass[aps,prd,twocolumn,superscriptaddress,nofootinbib]{revtex4-1}

\usepackage{amsmath,amssymb,amsfonts,bm,slashed,graphicx,longtable,color,array,multirow}
\usepackage{multirow, hyperref}
\usepackage{breakurl}

\newcommand{\gd}{\gamma_{\text{d}}}
\newcommand{\X}{\varphi}

\allowdisplaybreaks

\makeatletter
\g@addto@macro\bfseries{\boldmath}
\makeatother

\graphicspath{{./figures/}}

\newcommand{\expname}{CODEX-b}

\begin{document}

\title{Searching for Long-lived Particles: A Compact Detector for Exotics at LHCb}

\author{Vladimir V. Gligorov}
\affiliation{LPNHE, Universit\'{e} Pierre et Marie Curie, Universit\'{e} Paris Diderot, CNRS/IN2P3, Paris, France}

\author{Simon Knapen}
\affiliation{Ernest Orlando Lawrence Berkeley National Laboratory, University of California, Berkeley, CA 94720, USA}
\affiliation{Department of Physics, University of California, Berkeley, CA 94720, USA}

\author{Michele Papucci}
\affiliation{Ernest Orlando Lawrence Berkeley National Laboratory, University of California, Berkeley, CA 94720, USA}
\affiliation{Department of Physics, University of California, Berkeley, CA 94720, USA}

\author{Dean J. Robinson}
\affiliation{Physics Department, University of Cincinnati, Cincinnati OH 45221, USA}

\begin{abstract}
We advocate for the construction of a new detector element at the LHCb experiment, designed to search for displaced decays of beyond standard model long-lived particles, taking
advantage of a large shielded space in the LHCb cavern that is expected to soon become available. 
We discuss the general features and putative capabilities of such an experiment, as well as its various advantages and complementarities with respect to the existing LHC experiments and proposals such as SHiP and MATHUSLA. For two well-motivated beyond Standard Model benchmark scenarios -- Higgs decay 
to dark photons and $B$ meson decays via a Higgs mixing portal -- the reach either complements or exceeds that predicted for other LHC experiments.
\end{abstract}

\maketitle

\section{Introduction}
Deep and long-standing questions concerning the constitution of Nature remain unanswered. These include the (un)naturalness and structure of the electroweak vacuum, as well as the 
origins of dark matter, the baryon asymmetry, neutrino masses, flavor hierarchies and CP violation. Well-motivated theoretical frameworks that attempt to address these questions often 
predict the existence of metastable states, also known as long-lived particles (LLPs): LLPs with lifetimes up to the sub-second regime are broadly consistent with cosmological bounds, 
opening a large parameter space to be explored.

Many extensions of the Standard Model motivated by the hierarchy problem predict LLPs: LLPs have been studied in the context of gauge-mediated 
supersymmetry~\cite{Dimopoulos:1996vz,Liu:2015bma}, R-parity violating supersymmetry~\cite{Hall:1983id,Aulakh:1982yn,Liu:2015bma,Csaki:2015uza}, stealth supersymmetry~\cite{Fan:2011yu}, 
mini-split supersymmetry~\cite{Arvanitaki:2012ps,ArkaniHamed:2012gw}, neutral naturalness~\cite{Chacko:2005pe,Burdman:2006tz,Cai:2008au} and certain relaxion models~\cite{Graham:2015cka,Flacke:2016szy}. An extensive literature contemplates LLPs in 
models of dark matter~\cite{Baumgart:2009tn,Kaplan:2009ag,Dienes:2012yz,Kim:2013ivd,Falkowski:2014sma,Hochberg:2015vrg,Co:2015pka,Davoli:2017swj}, baryogenesis~\cite{Cui:2012jh,Cui:2014twa,Barry:2013nva}, 
neutrino masses~\cite{Graesser:2007yj,Graesser:2007pc,Helo:2013esa,PhysRevLett.115.081802,Batell:2016zod,Antusch:2016vyf,Dev:2017dui} and 
hidden valleys~\cite{Strassler:2006im,Han:2007ae,Strassler:2006ri,Strassler:2006ri}, predicting a wide range of LLP production and decay morphologies. 
Finally, the discovery of the Higgs boson~\cite{Chatrchyan:2012xdj,Aad:2012tfa} opens up the possibility of Higgs mixing portals that generically admit 
exotic Higgs decays into LLPs. 

ATLAS, CMS and LHCb have developed numerous programs to search for LLPs. Because of their large geometric acceptance and integrated luminosity, ATLAS and CMS will typically provide
the best reach for charged, colored or relatively heavy LLPs (see e.g.~\cite{CMS-PAS-EXO-16-036,Aaboud:2016dgf,CMS-PAS-EXO-16-003,ATLAS-CONF-2016-103}). Softer and/or short-lived 
final states tend to be more problematic due to triggering challenges and the large irreducible backgrounds inherent to high luminosity hadron collisions. LHCb, by contrast, may search 
for soft $\mathcal{O}(\text{GeV})$ final states, so long as the decay occurs within the Vertex Locator (VELO) system~\cite{Aaij:2016qsm,Aaij:2015tna,Pierce:2017taw,Antusch:2017hhu}. LHCb also adds sensitivity to states 
with a combination of high mass and short decay time~\cite{Aaij:2017mic,Aaij:2016xmb,Aaij:2016isa,Aaij:2015ica,Aaij:2014nma}. Longer-lived, GeV-scale LLPs
are difficult for all three experiments, but are theoretically well-motivated by various models of, e.g., asymmetric dark matter~\cite{Kaplan:2009ag,Kim:2013ivd}, strongly-interacting 
dark matter~\cite{Hochberg:2014dra,Hochberg:2015vrg} and neutral naturalness~\cite{Craig:2015pha}. Some scenarios may be searched for with current and future beam dump experiments such
as NA62~\cite{NA62:2017rwk} or SHiP~\cite{Alekhin:2015byh}, but comprehensive coverage should include a large sample of Higgs bosons, which at the moment can only be supplied by the LHC. 

In this paper we propose to take advantage of a large shielded space in the LHCb cavern that is expected to  become available after the pre-Run 3 upgrade, to construct a Compact 
Detector for Exotics at LHCb (``\expname{}''). Apart from generic LLP searches, by virtue of its location, such a detector is well-suited to probe for GeV-scale LLPs, for instance 
generated by Higgs mixing or dark photon portals. With a modest amount of additional shielding from the primary interaction point, \expname{} can operate in a low background environment, 
eliminating the triggering challenges associated with ATLAS and CMS. The modest size of \expname{} is anticipated to translate to relatively low construction and maintenance costs and a 
relatively short construction timescale with proven, off-the-shelf components. It should also be emphasized that \expname{} may provide complementary data, at relatively low cost, to potential 
discoveries in other proposed or existing experiments. 

We are aware of several other recent proposals in the same spirit. The most directly comparable is the MATHUSLA proposal~\cite{Chou:2016lxi}, 
which intends to operate at the surface above ATLAS or CMS, using an air-filled fiducial volume equipped with multiple horizontal tracking surfaces, and which may in some cases probe up to $\mathcal{O}(1\,\text{s})$ LLP lifetimes.  A recent analysis of the various key features and capabilities 
for this LLP search, using a template of tracking and partial reconstruction techniques, and an in-depth analysis of relevant backgrounds, can be found in Refs~\cite{Curtin:2017izq,Chou:2016lxi}. \expname{} necessarily has various commonalities with this proposal, though the backgrounds and configuration we examine here both differ in several critical aspects, and the theoretical prospects may also differ depending 
on the ultimate configuration and technologies chosen. Other planned or proposed LLP detectors located near the LHC experiments, include: milliQan~\cite{Haas:2014dda,Ball:2016zrp} which makes use of a drainage gallery above the CMS interaction point and is intended specifically to search for millicharged particles; and FASER~\cite{Feng:2017uoz}, which intends to operate a few hundred 
meters downstream of the ATLAS or CMS interaction point, looking for forward-produced light weakly-coupled particles. 

In what follows we describe the general features and putative capabilities of \expname{} and estimate the reach by focusing on two benchmark models, that illustrate the complementarity and 
relative advantages of \expname{} compared to other LHC experiments. These include: (i) A spin-1 massive gauge boson, $\gd$, that is produced through the exotic Higgs decay $h\rightarrow\gd\gd$.
 This dark photon can subsequently decay to charged Standard Model fermions through mixing with the photon~\cite{Schabinger:2005ei,Gopalakrishna:2008dv,Curtin:2014cca,Strassler:2008bv}.  
(ii) A light scalar field, $\X$, that mixes with the SM Higgs boson. Such a particle can be abundantly produced in 
inclusive $b \rightarrow s \X$ decays~\cite{Willey:1982dk,Chivukula:1988lo,Grinstein:1988yu}, subsequently decaying back to SM fermions through the same Higgs portal. 
We further study a proof-of-concept implementation of tracking, in order to verify that reconstruction of the decay vertices can be done efficiently, and 
explore the capabilities to reconstruct the exotic state mass using timing information. Details of the commonalities and complementarities between \expname{} and other proposed LLP searches will also be discussed.

\section{General strategy}

\subsection{Layout and capabilities}
\begin{figure*}[t]
	\includegraphics[width = 0.75\linewidth]{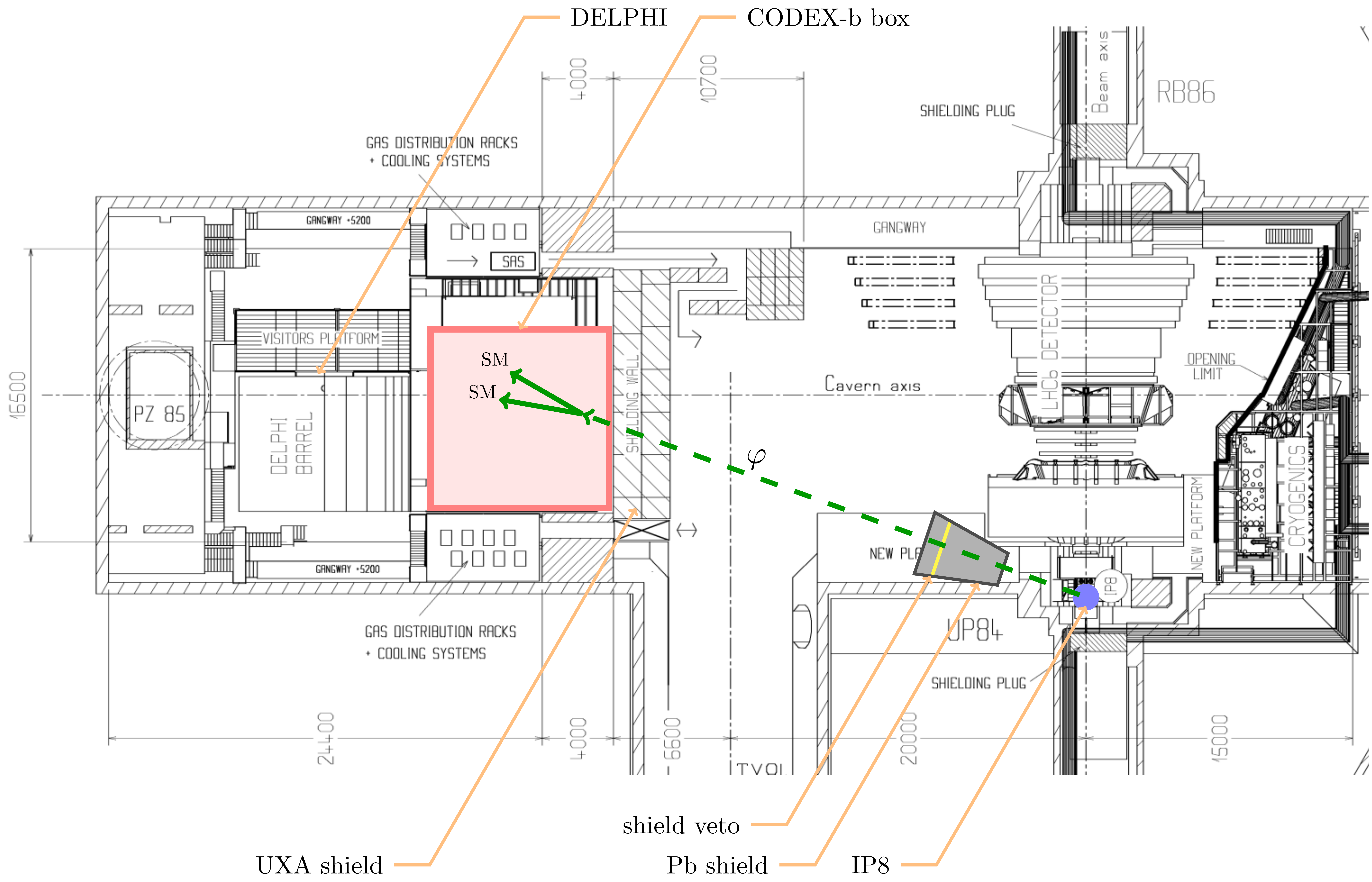}
	\caption{Layout of the LHCb experimental cavern UX85 at point 8 of the LHC~\cite{cavern}, overlaid with the \expname{} apparatus.} 
	\label{fig:LHCbCav}
\end{figure*}

The LHCb experiment is a single arm forward spectrometer, located at interaction point 8 (IP8) on the LHC beam line, previously occupied by the DELPHI experiment at LEP.   Apart from LHCb itself, key features for this discussion are the data acquisition (DAQ) space and DELPHI barrel exhibit, both located behind a 3\,m thick concrete radiation shield in the UXA cavern, approximately $25$\,m from IP8. The planned relocation of the DAQ to the surface during the upcoming pre-Run 3 upgrade~\cite{Bediaga:1443882} leaves available a ready-made underground cavity, which is large enough to house a detector with fiducial volume $10\times 10 \times 10$\,m in size, or  $20 \times 10 \times 10$\,m should DELPHI also be removed.\footnote{The old DAQ area also housed the low and high voltage power supplies of the different subdetectors, which would then have to move elsewhere in the cavern, for example to what is now the DELPHI public viewing gallery. Depending on the required infrastructure in UXA, it may therefore actually be more feasible to remove DELPHI and house the detector volume in its place, but such optimization is beyond the scope of this paper.} We shall, for the sake of brevity, refer to the fiducial volume as `the box'. For all sensitivity projections we use a $10\times 10 \times 10$\,m fiducial volume located as close as possible to the IP, specifically the coordinates of the fiducial volume are from $x=26$ to $36$\,m (transverse), $y=-7$ to $3$\,m (vertical) and $z = 5$ to $15$\,m (forward). The current layout of the LHCb cavern is shown in Fig.~\ref{fig:LHCbCav}, with the potential \expname{} location overlaid in red.

Our proposal, in short, is to instrument this space with tracking layers, in order to search for decays-in-flight of LLPs generated at IP8. (Additional shielding near the IP will be required to suppress backgrounds, discussed in detail below.) To our knowledge, no space of this size is available near the ATLAS and CMS interactions points; for example the drainage gallery that is to house MilliQan is only $\sim 3$\,m in diameter~\cite{Ball:2016zrp}. 

At the very least, in this type of experimental setup (cf. e.g. the MATHUSLA proposal~\cite{Chou:2016lxi}) one may search for at least two tracks originating from the same reconstructed vertex: A signal of the LLP decay $\varphi \to f f$ or $\varphi \to f f X$, where $\varphi$ is the LLP, $f$ is any trackable SM final state, and $X$ is anything else. Apart from $2$-body decays to SM particles, this morphology can include decays with additional invisible particles, e.g. $\varphi \to f f \varphi'$ or semileptonic $\varphi \to f \ell \nu$. The former are of interest for (supersymmetric) theories with squeezed spectra, in which the LLP may be an excited state with a small mass splitting from a lighter stable state. The signature of such multibody decays is acoplanarity of the tracks with respect to the decay vertex displacement from the IP. Multibody signatures may be more difficult to reconstruct, for example if the tracking efficiency drops sharply at low momentum due to multiple scattering. On the other hand, if these lower momentum tracks can be reconstructed, we may be able to determine their momentum with reasonable accuracy using time-of-flight information (see also Sec.~\ref{sec:massmeas}). A multitrack decay signature consistent in time and space with a mother particle originating from an LHC collision and inconsistent with the mass of a known SM particle would therefore be a very compelling sign of new physics (NP).

As we explore in some detail below, the relative proximity of the box to the interaction point, and the experimentally clean environment, produces a new physics reach that may complement or exceed those of other LHC experiments, despite the lower luminosity at the LHCb interaction point. Moreover, this setup may feature various other possibilities and capabilities:
\begin{enumerate}
	\item[(i)] A large amount of empty space between IP8 and the radiation shield can also be exploited if the LLP has significant branching fraction to muons, that typically punch through the UXA radiation shield.
	Here the signal comprises two muons whose tracks reconstruct a decay vertex in the region before the box itself. We call this region the `muon shadow'. Vertex reconstruction and control of backgrounds inside the muon shadow may be challenging, so we do not focus on this option hereafter, except for benchmarks with relatively hard muon final states.
	\item[(ii)] The proximity of the box to LHCb -- approximately only four bunch crossing times for relativistic objects -- may permit it to interface with LHCb's planned triggerless readout, allowing for identification and at least partial reconstruction of the LLP event. For the benchmarks we consider here, this may enable one to tag a VBF jet for Higgs decays, or an associated $K^{(\ast)}$ for $B$ decays.
	\item[(iii)] The modest size of the fiducial volume may also permit, in principle, implementation of more ambitious detection technologies such as calorimetry or time-of-flight, providing momentum reconstruction and particle identification that will aid in the confirmation of a discovery.
\end{enumerate}

\subsection{Reach intuition}
The geometric acceptance of the \expname{} box is~$\sim 1\%$ (normalized to $4\pi$). The LLP reach is attenuated further by the distribution of the LLP production and interplay between the LLP lifetime $\tau$ and the box depth. The number of LLP decay vertices expected in the box
\begin{equation}
	N_{\text{box}}  =  \mathcal{L}_{\text{LHCb}}\times \sigma_{pp \to \varphi X} \times \int_{\text{vol}}\!\! \frac{d \varepsilon (r,\eta)}{dV}\,dV\,,
\end{equation}
where the location of the box is specified by an azimuthal angle, the distance from the IP, $r$, and the pseudorapidity, $\eta$. In these coordinates, the differential fiducial efficiency is
\begin{equation}
	 \frac{d \varepsilon (r,\eta)}{dV} =  \frac{ 1}{2\pi r^2 c\tau}  \int\!\!d \beta\; w(\beta,\eta)\times\frac{e^{-r/(c\tau \beta\gamma)}}{\beta\gamma}   .
\end{equation}
with $\gamma$ and $\beta$ the usual kinematic variables. The function $w(\beta,\eta)$ is the differential probability of producing the LLP with pseudorapidity $\eta$ and velocity $\beta$, and is typically obtained from Monte Carlo.

To gain a rough sense of the achievable fiducial efficiency, let us assume $w$ is factorizable into a $\delta$-function in $\beta$ at $\beta_0\gamma_0 \sim 3$ and a flat distribution in pseudorapidity for $|\eta| < \eta_0 \sim 5$. This is a reasonable approximation for, e.g., an exotic Higgs decay. 
That is, $w(\beta,\eta)  \approx \delta(\beta-\beta_0)/(2\eta_0)$ on the box domain $\eta \in [0.2,0.6]$. The fiducial efficiency is then approximately
\begin{equation}
	\varepsilon_{\text{box}} \simeq \frac{0.4}{2\eta_0}\frac{|\phi_2 - \phi_1|}{2\pi} \Big[e^{-r_1/r_0} - e^{-r_2/r_0} \Big]\,,
\end{equation}
with $r_0 = c\tau \beta_0\gamma_0$. Using $|\phi_2 -\phi_1| \sim 10/25$, $r_1 \sim 25\,\text{m}$, $r_2 \sim 35\,\text{m}$, one estimates a maximum fiducial efficiency $\varepsilon_{\text{box}} \sim 10^{-3}$. In the long (short) lifetime regime $c\tau \gg r_{1,2}$ ($c \tau \ll r_{1,2})$, this efficiency is linearly (exponentially) suppressed by  $|r_2 - r_1|/r_0$ ($e^{-r_1/r_0}$). In the case of Higgs decay to dark photons, e.g., this translates to a maximal $2\sigma$ exclusion reach of $\text{Br}[h \to 2\gd] \sim 10^{-4}$, for $\mathcal{L} =300$\,fb$^{-1}$ expected after Run 5. We confirm this estimate with a more detailed simulation below.

\subsection{Tracking}
\label{sec:tracking}
In order to demonstrate the feasibility of the proposed detector, we have studied a simple tracking layout
based on RPC strip modules with 1\,cm$^{2}$ effective granularity. Such modules typically also have 1\,ns or better timing resolution,
which may be useful for background rejection or improving the reconstruction of slow-moving signals. 

As there is no magnetic field to provide a momentum estimate, it is crucial that we have the best possible spatial
resolution on the decay vertex of our hypothetical long-lived particle. The most important parameter for vertex
resolution is the distance between the vertex and the first measured point on the decay product trajectories. Further, softer LLPs may have large opening angles, so that their decay products can exit the box through any of its six faces with non-negligible probability. For these reasons, we instrument each of the six box faces with a sextet of RPC layers, and we also add five equally-spaced triplets of RPC layers along the depth of the box (i.e. $y$-$z$ aligned planes with transverse spacing $\delta x = 1.67$\,m). The layers in each RPC sextet or triplet are spaced at 4\,cm intervals.
For the purpose of our estimates of the efficiency, we consider an event to be reconstructed if (i) each track has at least 6 hits, (ii) each track has at least $600$\,MeV of momentum (cf. Ref.~\cite{LHCbtwiki:2015}) and (iii) none of the hits are shared between the tracks. A detailed study of tracking performance for different particle species, 
including the effects of multiple scattering, is left for a future study. 

We study the vertex reconstruction efficiency achieved by the above-specified layout for the $h\rightarrow \gd\gd$ and $B\to \X X_s$ benchmarks over a range of $\gd,\X$ masses and lifetimes. Details of these models and the reach analysis are provided below in Sec.~\ref{sec:reach}. 
For the purposes of this study we decay the long-lived particle into two charged tracks, not least because a realistic estimate of the performance for multi-body decays will depend much
more strongly on assumptions about the detector layout and the impact of multiple scattering.  
The reconstruction efficiencies for the benchmark scenarios are shown in Tab.~\ref{tab:effsummary}. 
In the case of $B\to \X X_s$ decays, the efficiency is dominated by the momenta of the decay products. 
By contrast, the main source of inefficiency for $h \to \gd\gd$ decays, particularly at low $\X$ mass, is the small opening
angle of the decay, which results in partially overlapping decay products. We currently reject all signal decays for which the
first measured points of the decay products are within the nominal detector granularity (1\,cm) of each other, however this inefficiency may be reducible by
improving the detector granularity or by allowing fewer than six hits on a track. 

\begin{table}[t]
\centering
\renewcommand*{\arraystretch}{1.3}
\newcolumntype{C}{ >{\centering\arraybackslash} m{0.7cm} <{}}
\begin{tabular*}{\linewidth}{@{\extracolsep{\fill}}l|CCC|CCCCC}
\hline
$c\tau$ (m) & \multicolumn{3}{c|}{$m_\X ~[B \to X_s\X]$} &  \multicolumn{5}{c}{$m_{\gd}~[h \to \gd\gd]$} \\ 
     & $0.5$ &  $1.0$ & $2.0$ & $0.5$ & $1.2$ & $5.0$ & $10.0$ & $20.0$ \\ \hline\hline
0.05  & --   & --   & --   & 0.39 & 0.48 & 0.50 & --   & --     \\
0.1   & --   & --   & --   & 0.48 & 0.63 & 0.73 & 0.14 & --     \\
1.0   & 0.71 & 0.74 & 0.83 & 0.59 & 0.75 & 0.82 & 0.84 & 0.86   \\
5.0   & 0.55 & 0.64 & 0.75 & 0.60 & 0.76 & 0.83 & 0.86 & 0.88   \\
10.0  & 0.49 & 0.58 & 0.74 & 0.59 & 0.75 & 0.84 & 0.86 & 0.88   \\
50.0  & 0.38 & 0.48 & 0.74 & 0.57 & 0.75 & 0.82 & 0.87 & 0.88   \\
100.0 & 0.39 & 0.45 & 0.73 & 0.62 & 0.77 & 0.83 & 0.87 & 0.89   \\
500.0 & 0.33 & 0.40 & 0.75 & --   & --   & --   & --   & --     \\
\hline
\end{tabular*}
\caption{Efficiency of reconstructing at least two tracks for a LLP decaying in the fiducial volume, for both $B\to  X_s\X$ and $h\to \gd\gd$ scenarios, for various lifetimes. Masses are in GeV.}
\label{tab:effsummary}
\end{table}

\subsection{Backgrounds and shielding\label{sec:BG}}
Backgrounds from cosmics are suppressed because of the shielded underground location, by the fact that most cosmics are out of time with respect to the LHC collisions,
and because the signal track velocities should have a significant horizontal component, as the IP is horizontally displaced from the detector. The rate of cosmic muons incident on CMS was simulated to be $\sim$1 kHz~\cite{Drollinger:2005dm}, the vast majority of which would traverse the box from the top to bottom. Once timing and directionality are accounted for, this background is therefore subdominant to muons from the primary vertex, which have a rate of the order 1-10 kHz~\cite{ATL-DAQ-PUB-2016-001}. 

Apart from the existing UXA radiation shield, additional shielding is required to attenuate collision backgrounds to manageable levels. Prompt muons, or muons from decays-in-flight of charged pions and kaons, may typically generate tracks in the detector volume, but the likelihood of two such muon tracks themselves reconstructing a fake vertex inside the box is expected to be negligible once muons that point back to vicinity of the IP are vetoed.  However, muons may scatter on air inside the detector volume, producing tracks either from ionized electrons, elastic pair production or muon-nucleon deep inelastic scattering (see Ref.~\cite{Groom:2017ms} for a review). Other backgrounds arise from neutral kaon decays producing track pairs, and neutrons or neutrinos producing scattering events with morphologies resembling a signal. On top of the primary fluxes of leptons and hadrons, also secondary production in the shield itself can be significant, such that the inclusion of active veto components in the lead shielding is required. A ``shield veto'' for charged particles (mainly $\mu$'s and $\pi^{\pm}$) would remove events where, for example, a muon produces secondary neutrons or kaons in the last few interactions lengths of the concrete shield, that then enter the box and either scatter on nuclei or decay, respectively.

The minimum amount of shielding can be estimated by sufficiently reducing the prompt $K_L$ and neutron components. From a simulated minimum bias sample generated with \texttt{Pythia 8}~\cite{Sjostrand:2006za,Sjostrand:2014zea}, the fiducial efficiency for promptly-produced $K_L$'s and $n$ to enter the box is $\varepsilon_{\text{box}} \simeq 4.8\times 10^{-3}$ and $5.1\times 10^{-3}$ respectively. For a QCD cross-section $\sigma_{\text{qcd}} \simeq 100$~mb, at $300$\,fb$^{-1}$ of integrated luminosity one naively requires $\simeq 32\lambda$ of shielding -- $\lambda$ is an interaction length -- to suppress these backgrounds to $\lesssim 1$ event. The natural location for such a shield is as close to the IP as possible, as this reduces the volume of shielding material required. Based on the cavern geometry shown in Fig.~\ref{fig:LHCbCav}, we focus on the case that a Pb shield is installed covering the detector geometric acceptance, beginning approximately $5$m from the IP and several meters in depth. (One may also consider other shielding materials such as tungsten, but the qualitative conclusions remain the same.) As a proof-of-concept, based on the quoted PDG nuclear interaction lengths for energetic neutrons, respectively $18$\,cm and $42$\,cm~\cite{PDG:2016}, we further focus on the following shielding configuration: $3.6$m of Pb, corresponding to $20\lambda$; a shield veto comprising RPC or scintillator layers with efficiency $\varepsilon_{\text{veto}}$; an additional $0.9$m of Pb, corresponding to $5\lambda$; and finally the UXA concrete shield, comprising approximately $7\lambda$. We refer to this as a ``(20+5)$\lambda$'' Pb shielding configuration and consider it as our benchmark in the following. Further optimization of the shield design is beyond the scope of this study.

We model the shielding response to backgrounds -- propagation of primary muons, pions, kaons, neutrons and protons and secondary production of these and neutrinos and photons -- with a preliminary \texttt{Geant4}~\cite{Agostinelli:2002hh,Allison:2006ve,Asai:2015xno} simulation which includes propagation in lead, air and concrete. We simulate the initial muon flux for $ E_\mu \lesssim 7$\,GeV with \texttt{Pythia 8};  pion, kaon and heavy flavor decays in minimum bias events produce the dominant primary contribution in this regime.  Based on the geometry of Fig.~\ref{fig:LHCbCav}, we conservatively include secondary muons produced by pions or kaons that decay within $5$\,m transverse of the IP, and within $3$\,m forward of the IP. The resulting estimated muon flux for $E_\mu \gtrsim 3$\,GeV is typically $\mathcal{O}(\text{few})$ larger than a naive estimation of the same from inclusive muon spectra measured at ATLAS~\cite{ATLAS-CONF-2010-035}.  To estimate the spectrum for $E_{\mu} > 7$\,GeV, we use the shape of measured muon inclusive spectra at ATLAS~\cite{Aad:2011rr} and rescale them to match onto our simulated results, assuming the muon spectrum is roughly flat in the $|\eta| < 2$ regime and modified by only $\mathcal{O}(1)$ factors upon mapping from $7$ to $14$ TeV, the latter verified using \texttt{FONLL}~\cite{Cacciari:1998it,Cacciari:2001td}. Primary fluxes and spectra of kaons, pions, protons and neutrons in the box acceptance are generated from a minimum bias sample produced with \texttt{Pythia 8}.

The simulation of the (20+5)$\lambda$ shielding response proceeds as follows: All primary fluxes are propagated through $20\lambda$ of Pb up to the shield veto, then partitioned into intermediate charged and neutral fluxes, since only charged fluxes can be rejected by the active veto. We then propagate these charged and neutral fluxes through the remaining lead, 
air and concrete layers separately and refer to them as the reducible and irreducible fluxes respectively:
The veto has power to reject secondaries produced from charged states downstream  -- in the matter between the veto and the detector -- but not 
neutral secondaries produced upstream -- in the matter between the IP and the veto -- or produced by neutral fluxes. 

\begin{figure}[ht]
	\includegraphics[width = 0.8\linewidth]{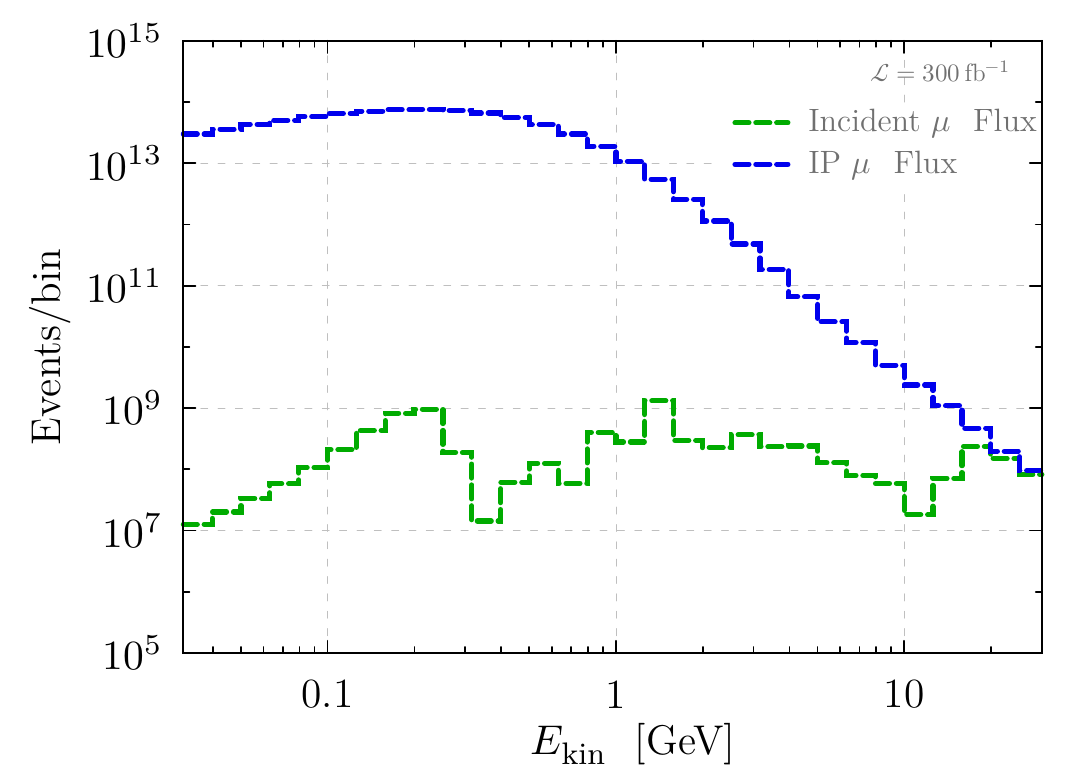}
	\includegraphics[width = 0.8\linewidth]{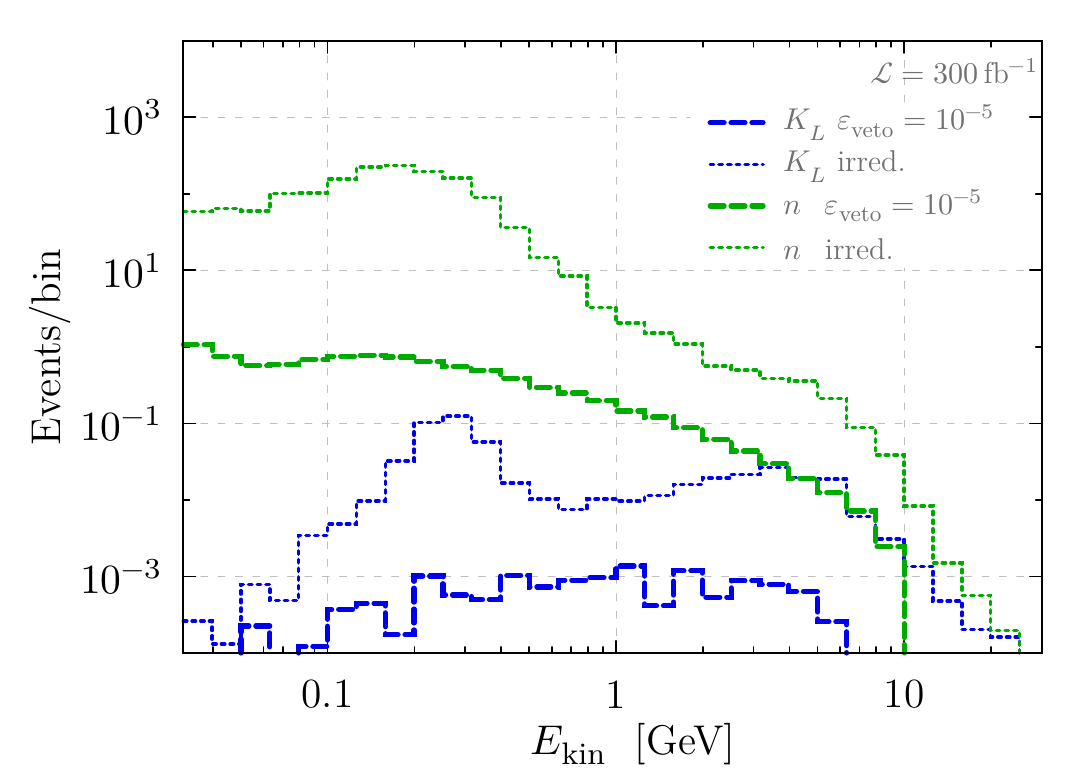}
	\caption{(Top) Energy dependence of muon number flux before propagation through shielding (blue), and of muon number flux incident on the fiducial volume (green) after propagation through shielding. (Bottom) Energy dependence of neutral kaon (blue) and neutron (green) shield vetoable and irreducible fluxes. The vetoable fluxes are specified with a rejection factor $\varepsilon_{\text{veto}} = 10^{-5}$ already applied.} 
	\label{fig:MuE}
\end{figure}

In Fig.~\ref{fig:MuE} we show the kinetic energy distributions for the flux of muons, neutral kaons and neutrons entering the box, after propagating through the shield. Both neutrons and kaons are quite soft, so that only a small fraction are kinematically capable of producing multiple tracks and are sufficiently boosted to do so within timing constraints. In Table~\ref{tab:bkg-yields} we show the results of this preliminary analysis, including approximate energy thresholds to produce charged tracks and/or satisfy timing constraints. The dominant contributions in both shield reducible and irreducible populations arise from neutrons. (The neutrinos have a sub-pb cross-section, resulting in $\ll1 $ scattering event for a population of $\mathcal{O}(10^6)$ size.) The neutron incoherent cross section on nitrogen $\sim 1$b, so that the probability of a neutron scattering on air somewhere in the box is $\sim 5\%$. From Table~\ref{tab:bkg-yields}, the number of neutron scattering background events is then $\sim 0.4$. Moreover, if needed, any surviving unvetoable neutrons can be further suppressed with a marginal increase in the Pb shield, or based on the event morphology given their softness. The vetoable neutron population suggests $\varepsilon_{\text{veto}} \sim 10^{-5}$ is required, which is used in Fig.~\ref{fig:MuE}~(bottom) for the neutron and $K_L$ fluxes. These results broadly agree with a simplified propagation model making use of the muon average energy loss $\langle -d E_\mu/dx \rangle$ for Pb and standard concrete, taken from PDG data~\cite{PDG:2016}, the quoted neutron interaction lengths~\cite{PDG:2016}, the measured neutral kaon absorption cross-section in Pb at low energies~\cite{LAKIN1970677,PhysRevLett.42.9} and the inclusive muoproduction cross-section for neutral kaons~\cite{PhysRevLett.40.1614,PhysRevLett.45.765,ARNEODO1984156,ENT1998385,Anelli:2015pba}.

\begin{table}[t]
\centering
\renewcommand*{\arraystretch}{1.3}
\begin{tabular*}{\linewidth}{@{\extracolsep{\fill}}l|c|c|c}
\hline
\multirow{4}{2cm}{BG species} & \multicolumn{2}{c|}{Particle yields} & \multirow{4}{*}{Baseline Cuts}  \\ \cline{2-3}
 & \multirow{3}{2cm}{\centering irreducible by shield veto} & \multirow{3}{2cm}{\centering reducible by shield veto} & \\
 & &  & \\
 & &  & \\
 \hline \hline
 $n+\bar n$ & 7 & $5 \cdot 10^4$  &$E_{\text{kin}} > 1\,\text{GeV}$ \\
 $K^0_L$ & 0.2 & $9 \cdot 10^2$  &$E_{\text{kin}} > 0.5\,\text{GeV}$ \\
 $\pi^\pm+K^\pm$ & 0.5 & $3\cdot 10^4$ & $E_{\text{kin}} > 0.5\,\text{GeV}$ \\
 $\nu+\bar\nu$ & 0.5 & $2\cdot 10^6$ & $E > 0.5\,\text{GeV}$ \\
 \hline
\end{tabular*}
\caption{Particle yields in \expname{} from the preliminary \texttt{Geant4} background simulation. Yields are classified by particle species and calculated for $300\,\text{fb}^{-1}$ of integrated luminosity using the (20+5)$\lambda$ Pb shield. Separate yields for (ir)reducibility by the active shield veto are indicated. The irreducible $\pi^\pm$, $K^\pm$ yields can be vetoed by the box tracking layers themselves. The yield of $\sim 7$ irreducible neutrons corresponds to $\lesssim 0.4$ BG events, assuming the box volume is filled with air. The applied cuts conservatively select the kinematic thresholds necessary for producing charged tracks in the box.}
\label{tab:bkg-yields}
\end{table}

The muon flux entering the box is shown in Fig.~\ref{fig:MuE} (top). To conservatively account for dangerous muon-air interactions, we (over)estimate the cross-section via the total muon-proton cross-section with energy exchange $\delta E_\mu > 0.1$\,GeV. This cross-section includes contributions from: elastic pair-production, $\mu p \to \mu e^+ e^- p$; inelastic photonuclear processes, $\mu n/p \to \mu X$, with $X$ a hadronic final state containing at least one baryon; bremsstrahlung, for which the radiated photon has low, but non-zero conversion probability; and ionization. Between $10$ and $100$\,GeV, this cross-section is roughly constant, being $\mathcal{O}(\text{mb})$~\cite{Groom:2017ms, Bulmahn:2008fa}. Keeping only muons above $0.1$\,GeV and including a $10^{-5}$ rejection factor by the shield veto, the number of air scattering muons in the box is ${\cal O}(70)$ for $\mathcal{L} = 300$\,fb$^{-1}$. These events can be rejected by the RPC layers on the detector faces. 

Neutrino air-scattering inside the detector volume, arises from either the atmospheric $\nu$ flux or from the IP. For neutrinos with $E\sim 1$ GeV, the total neutrino-nucleon cross section is roughly $0.01\,\text{pb} \times (E_\nu/\text{GeV})$~\cite{Formaggio:2013kya}, which implies that the probability for a $\sim 1$ GeV neutrino to scatter in the fiducial volume is $\mathcal{O}(10^{-14})$. For an atmospheric neutrino flux of $\sim 1\, \text{cm}^{-1}\, E_\nu^{-1}\, s^{-1}$~\cite{Daum1995} , we expect the roughly $0.01$ collisions per year, which is revised to $\sim 0.05$ if the energy dependence of the cross section and neutrino flux are accounted for. If needed, this background can be reduced further by imposing timing and directionality constraints.

The neutrino background from the IP requires a proper simulation of the material budget in the LHCb cavern, which is beyond this proof-of-concept study. 
Once an energy threshold cut $E_\nu > 0.5$~GeV is included, the contribution to this flux from the decays of hadrons (mostly $\pi^\pm$) propagating through the shield is negligible, as indicated by Table~\ref{tab:bkg-yields}. Similarly, pions decaying in any dense material inside the cavern or in its walls overwhelmingly produce only soft neutrinos. Instead, the main contribution to the neutrino flux comes from pion decays in the empty spaces within the cavern. We conservatively estimate this flux by simulating the truth-level neutrino energy spectrum from pion and kaon decays with \texttt{Pythia 8}. We impose a $|\eta|<5$ cut and assume that the remaining neutrinos radiate out spherically from the IP. We further require that the decay lengths of the pion to be shorter than $20\,\text{m}$ in the forward direction, $z>0$, and require the neutrino energy to be larger than $0.5\,\text{GeV}$ as in Table~\ref{tab:bkg-yields}. (Harder neutrinos from heavy flavor decays have a flux at least $\mathcal{O}(10^{-2})$ smaller than in this estimate.) Folding this flux with the total neutrino-nucleon cross section~\cite{Formaggio:2013kya} to estimate an upper limit on the number of neutrinos scattering against air in the fiducial volume, we find roughly ${\cal O}(3)$ potential events, but we expect that this estimate is still very conservative. In particular, the neutrino cross section falls quadratically with decreasing energy below $1$\,GeV, no pion energy losses have been included, and the size of the empty cavern spaces have been overestimated. This conclusion is also supported by a more realistic simulation for the MATHUSLA proposal [54], which estimates at most $\mathcal{O}(3)$ IP generated neutrino events at the HL-LHC, but for $\sim 100$  times larger luminosity times geometric acceptance compared to \expname{}.

We finally note that the location of CODEX-b and the fact that all the relevant backgrounds originate in the LHC collisions provides a simple way of calibrating them: All that is required is to install 
a telescope consisting of a few RPC layers in the LHCb cavern. Using this one can measure background fluxes with and without the additional shielding proposed for
CODEX-b, and obtain a data-driven measurement of the background rates from which to extrapolate the expected numbers of background events in CODEX-b itself.

\section{Reach Analysis\label{sec:reach}}
\subsection{Exotic \texorpdfstring{$B$}{B} decays}
A Higgs mixing portal admits exotic inclusive $B \to X_s \X $ decays, in which $\X$ is a light CP-even scalar that mixes with the Higgs, with mixing angle $\theta \ll 1$. The inclusive branching ratio~\cite{Willey:1982dk,Chivukula:1988lo,Grinstein:1988yu}
\begin{equation}
	\label{eqn:BRIC}
	\frac{\text{Br}[B \to X_s \X]}{\text{Br}[B \to X_c e \nu]}  =  \frac{27 g^2 s_\theta^2}{256 \pi^2} \frac{m_t^4}{m_b^2 m_W^2} \frac{(1 - m^2_\varphi/m_b^2)^2}{f(m_c/m_b)} \bigg|\frac{V_{ts} V_{tb}}{V_{cb}}\bigg|^2\,,
\end{equation}
with $s_\theta\equiv\sin\theta$. This yields $\text{Br}[B \to X_s \X ] \simeq 6.~s_\theta^2(1 - m^2_\varphi/m_b^2)^2$ from the measured inclusive $B \to X_c e \nu$ semileptonic rate~\cite{PDG:2016}. This ratio is expected to break down for $m_\X$ approaching $m_b$.  Approximating the inclusive rate instead by adding the various exclusive contributions, we have checked that we do not recover sensitivity in this high mass region. In our simple benchmark model, the scalar $\X$ decays exclusively to SM  states, such that its lifetime is also controlled by $s_\theta^2$, though complicated by effects of hadronic resonances when $m_\X \sim 1$\,GeV, as well as threshold effects. The theory uncertainties in this region are substantial, and we use the data-driven result from~\cite{Fradette:2017sdd,Bezrukov:2009yw}. Another common choice is the perturbative spectator model~\cite{Gunion:1989we}. The difference in reach predicted by both models can be rather large, especially in the region around the masses of the $f_0$ mesons. As an example, for $m_\X=1$\,GeV the branching ratio is related to the lifetime via $\text{Br}[B \to X_s \X] (c\tau_\X / \text{1\,m} ) \sim 6 \times 10^{-10}$ in the data driven model, while the analogous number in the spectator model is $2\times 10^{-8}$.

We generate a $B$ meson sample with \texttt{Pythia 8}, enforcing the exclusive decay $B \to K \X$ as a proxy to estimate the box fiducial efficiency for $B \to X_s \X$.  We do not include muon shadow contributions, as the muon energy in these LLP decays is typically low, $E_\mu \lesssim \text{few}$\,GeV, so that muon penetration (scattering) through the concrete shield might be unacceptably low (high) for a decay vertex reconstruction. The peak box fiducial efficiency is $\sim 10^{-4}$ at $c \tau_\X \sim 10$\,m: The LLPs captured by the transversely located box are typically only mildly boosted.

\begin{figure}[t]
	\includegraphics[width =  \linewidth]{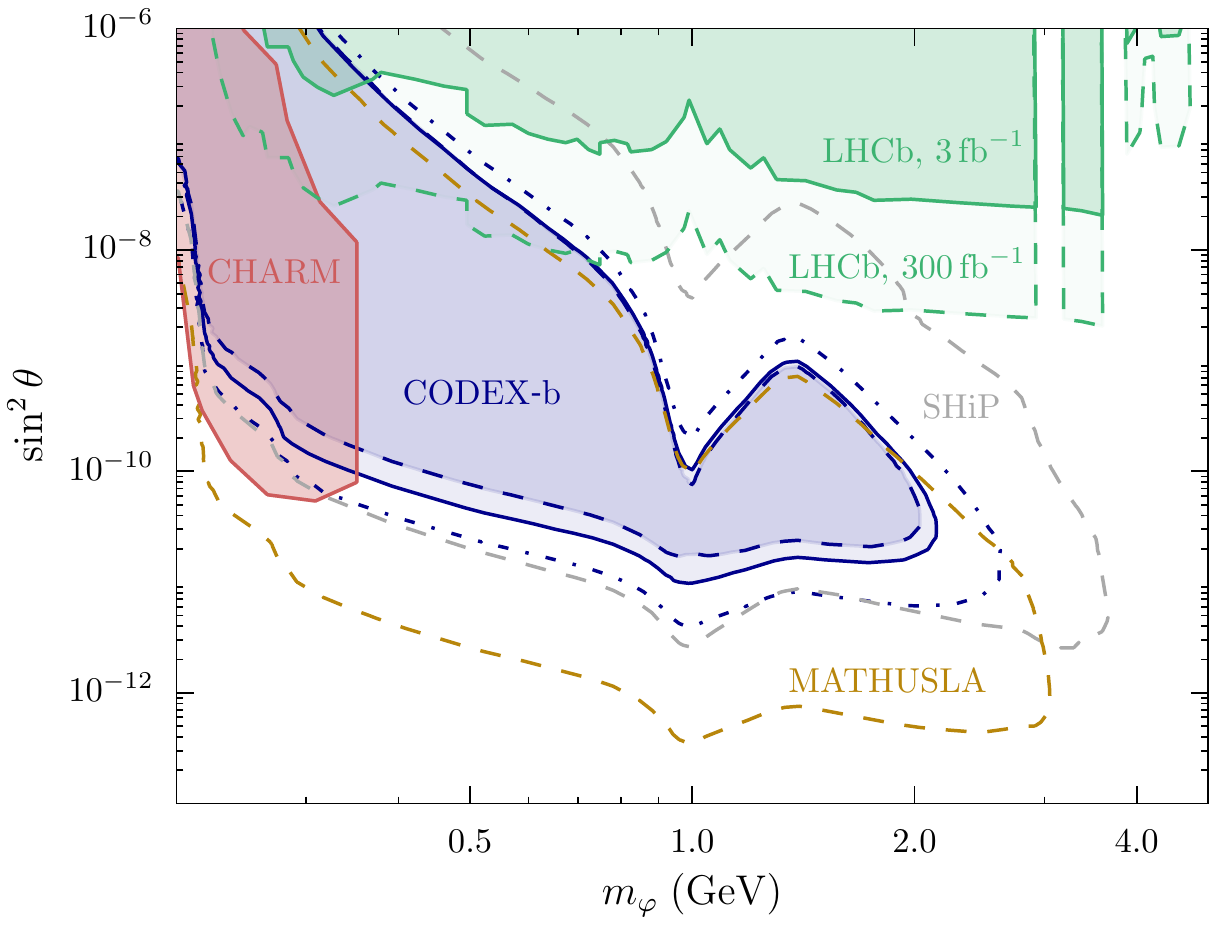}
	\caption{\expname{} reach for $B\rightarrow X_s\X$ in the $s^2_\theta$--$m_\X$ plane. Solid (dashed) blue line assumes 100\% (Tab.~\ref{tab:effsummary}) tracking efficiency. Dot-dashed line indicates the reach for $\mathcal{L} = 1\, \text{ab}^{-1}$.} 
	\label{fig:ThVM}
\end{figure}

In Fig.~\ref{fig:ThVM} we show the \expname{} reach (exclusion at greater than $95$\% CL) on the Higgs mixing portal $s^2_\theta$--$m_\X$ parameter space, compared with existing bounds from CHARM~\cite{BERGSMA1985458} and LHCb~\cite{Aaij:2016qsm}, as well as projected reaches for LHCb, MATHUSLA and SHiP. 
We assume a $b\bar{b}$ production cross-section of $500$\,$\mu$b. For the projected LHCb reach we rescaled the existing $B\rightarrow K (\X \rightarrow \mu\mu)$ limit~\cite{Aaij:2016qsm} under the (optimistic) assumption of zero background, implying that the limit on the fiducial rate scales linearly with the integrated luminosity. (A similar limit from $B\rightarrow K^\ast (\X \rightarrow \mu\mu)$ is slightly weaker~\cite{Aaij:2015tna}.) The sensitivity of MATHUSLA to this signature has concurrently been pointed out in \cite{Evans:2017lvd}. The curve here is our own recast of the MATHUSLA reach and agrees with the results in \cite{Evans:2017lvd}, up to small differences which can attributed to slightly different assumptions regarding the  width of $\X$. The original SHiP projection~\cite{Lanfranchi:2243034} was computed using a perturbative spectator model for the width of $\X$. To properly compare all experiments, the curve shown in Fig.~\ref{fig:ThVM} is a recast to the data-driven model in \cite{Fradette:2017sdd,Bezrukov:2009yw}, where we use the efficiency maps provided in~\cite{Lanfranchi:2243034}.

The lower extent of the reach in $s^2_\theta$ is determined by the total number of beauty hadrons and the \expname{} fiducial efficiency, while the upper extent of the $s^2_\theta$ reach is controlled by the $\X$ lifetime: A larger $s^2_\theta$ implies a larger rate of $\X$ production along with a shorter $\X$ lifetime, such that most $\X$'s decay before they reach the detector. One finds that \expname{} would significantly extend the reach of LHCb, and complement part of the projected parameter reach for SHiP as well as for MATHUSLA.

\begin{figure}[t]
	\includegraphics[width =  \linewidth]{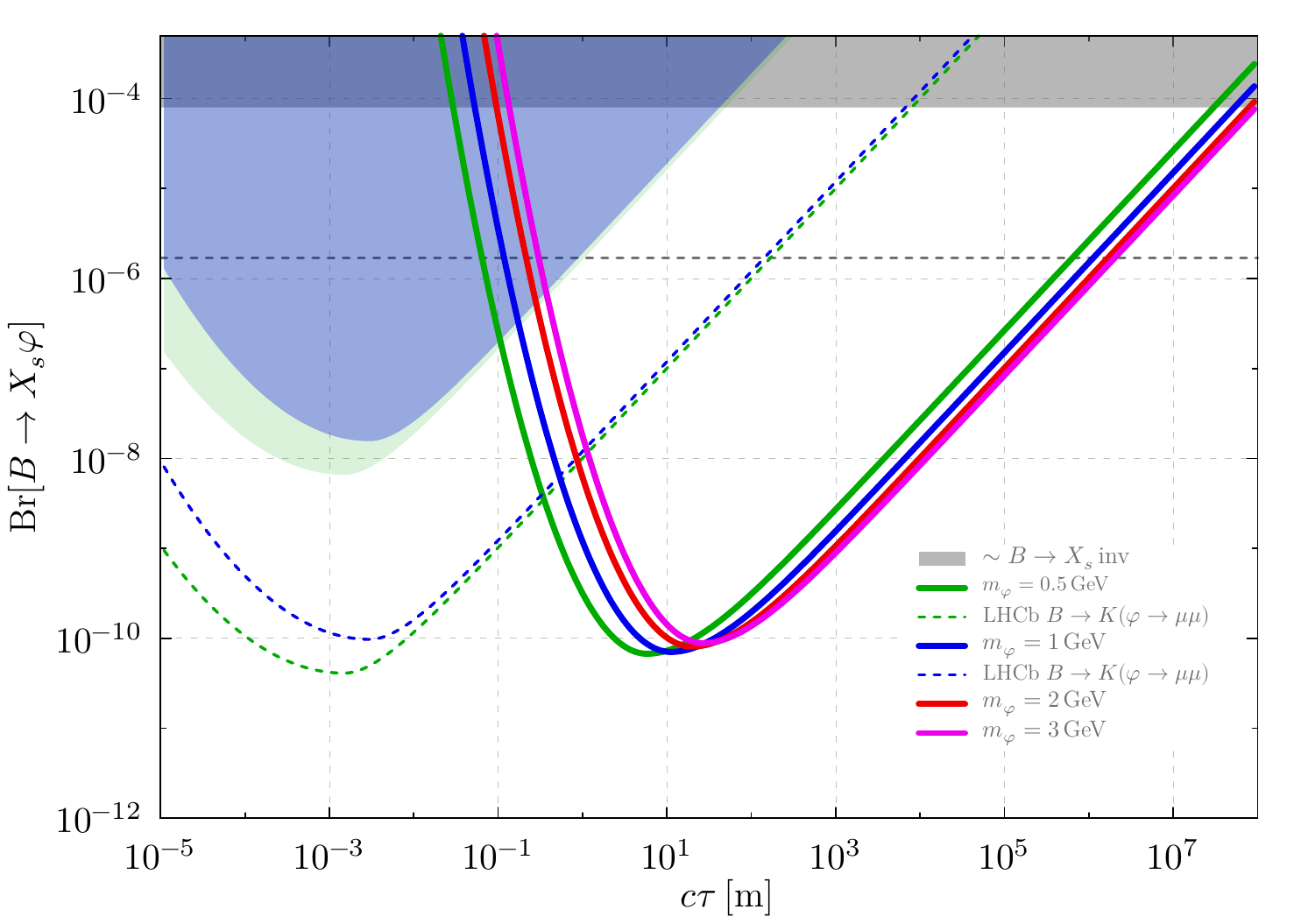}
	\caption{Inclusive \expname{} $B \to X_s \X$ reach (solid lines). The shaded regions (dashed lines) indicate current LHCb limits (300$\,$fb$^{-1}$ projection) from $B \to K(\X \to \mu\mu)$, rescaled to the inclusive process using the ratio of Eq.~\eqref{eqn:BRIC} and the theory predictions for the exclusive branching ratio~\cite{Batell:2009jf,Krnjaic:2015mbs}, and assuming $\text{Br}[\X \to \mu\mu] \simeq 30\%$  and $10\%$ for $m_\X = 0.5$~GeV and $1$\,GeV, respectively. Approximate current~\cite{PDG:2016} and Belle II projected~\cite{BelleIIreport} limits from $B \to K^{(*)}\nu\bar\nu$ precision measurements are also shown (gray shading and dashed line).}
	\label{fig:CTB}
\end{figure}

One may also consider more general portals that do not feature the fixed branching ratio-lifetime relations predicted by the simplest Higgs portal models. In Fig.~\ref{fig:CTB} we show the branching ratio reach for such theories, for various $\X$ mass benchmarks. Compared to LHCb, which searches for $B \to K(\X \to \mu\mu)$, a key advantage is that the reach is not sensitive to the model-dependent muonic branching ratio, only requiring instead that the final states are trackable. (Decays into neutral hadron pairs, such as $\pi^0 \pi^0$, cannot be seen without calorimetry, however such final states comprise at most $\mathcal{O}(30\%)$ of final states for $2m_\pi < m_\X \lesssim 1$\,GeV and typically otherwise comprise a much smaller contribution.) While the muon branching ratio is typically $\mathcal{O}(1)$ for $m_{\X} < 2m_K$ from kinematic considerations, at higher masses this branching ratio may drop precipitously to the sub-percent level. As an example, we show the projected LHCb reach in Fig.~\ref{fig:CTB} for $m_\X = 0.5$\,GeV compared to $m_\X =1$\,GeV.

\subsection{Exotic Higgs decays}
\begin{figure}[t]
	\includegraphics[width =  \linewidth]{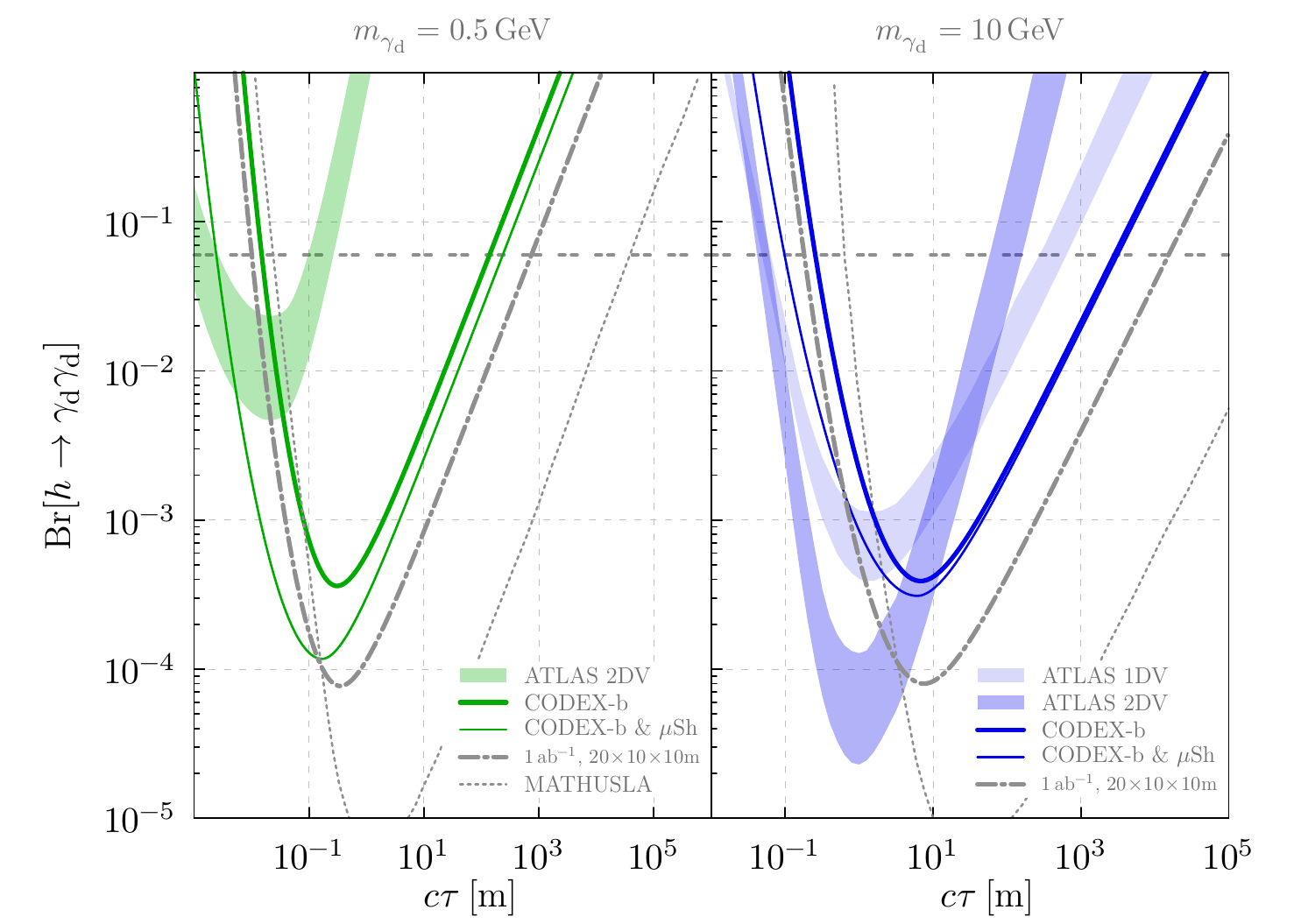}
	\caption{Higgs decay to dark photon reach, using the \expname{} fiducial volume alone and with the muon shadow `$\mu$Sh'. The $\gd \to \mu\mu$ branching ratio is taken from $e^+e^-$ data~\cite{Meade:2009rb}. Also shown is the \expname{} reach with $\mathcal{L} = 1$\,ab$^{-1}$ and a larger box, should DELPHI be removed. The approximate reach for MATHUSLA (gray dotted), rescaled from~\cite{Chou:2016lxi}, and $h \to \text{invisibles}$ is also shown (horizontal gray dashed)~\cite{CMS:2013xfa}. }
	\label{fig:HXX}
\end{figure}

Exotic Higgs decays to two dark photons may be generated by a kinetic mixing portal (e.g.~\cite{Schabinger:2005ei,Gopalakrishna:2008dv,Curtin:2014cca,Strassler:2008bv}). In the short lifetime limit, dark photons can be searched for with the main LHCb detector, in $D^\ast$ decays~\cite{Ilten:2015hya} or with an inclusive search~\cite{Ilten:2016tkc}. To estimate the \expname{} fiducial efficiency, we simulate gluon fusion Higgs production at IP8 with \texttt{Pythia 8}, with subsequent $h \to \gd\gd$ decay. The dark photon branching ratios to various SM final states are approximated from existing $e^+e^-$ data~\cite{Meade:2009rb}, which is relevant if one exploits the muon shadow. In Fig.~\ref{fig:HXX} we show the expected reach in $\text{Br}[h \to \gd\gd]$ for $m_{\gd} = 0.5$ and $10$ GeV benchmarks as a function of dark photon lifetime, for both the \expname{} fiducial volume and for the case that the muon shadow can be used.  For the $0.5$~GeV benchmark, the larger $\gd \to \mu\mu$ branching ratio enhances the reach of the muon shadow, compared to the $10$~GeV case.

A displaced vertex search at ATLAS/CMS has geometric acceptance $\sim 1$ (normalized to $4\pi$), and approximately $10$ times higher luminosity. Other than the trigger challenges associated with LLPs, a second crucial distinction is that the calorimeters comprise only $\sim 10\lambda$ of shielding compared to the $32\lambda$ shield in the \expname{} setup. Searches for light displaced objects in the ATLAS/CMS muon system are therefore expected to suffer from significant backgrounds from punch-through jets. To heavily reduce these backgrounds, it is often necessary to require two displaced objects, which is a significant penalty in reach for the long lifetime regime. We compare our $m_{\gd}=10$ GeV benchmark point with the projected sensitivity of searches for one and two displaced jets~\cite{Coccaro:2016lnz}. The latter estimate is based on the existing ATLAS displaced dijet search~\cite{Aad:2015uaa}. Its sensitivity deteriorates when the displaced vertices generate a low number of tracks, which occurs both for $m_{\gd}<10$ GeV and for models with small hadronic branching ratios. Neither difficulty applies to \expname{}. 

For our $m_{\gd}=0.5$ GeV benchmark, we compare with the ATLAS search for a pair of displaced lepton jets~\cite{ATLAS-CONF-2016-042}. This search is currently systematics limited, so that the range for our estimate of the HL-LHC ATLAS reach in the left panel of Fig.~\ref{fig:HXX} is bounded above by the current expected limit~\cite{ATLAS-CONF-2016-042} and bounded below by the current expected limit, rescaled under the assumption that the systematic uncertainties can be reduced with a factor of five. We expect backgrounds for a single displaced dilepton search would be prohibitively large. 

\subsection{Mass measurement\label{sec:massmeas}}
Aside from the discovery potential outlined above, \expname{} should also be capable of measuring the velocity of the LLP. 
For a given assumption on the production mechanism, this then allows for a mass measurement of the new state on a statistical basis. 
As is well known (see e.g. Ref.~\cite{Curtin:2017izq}), the geometry of two-body $\gd/\X$ decays to massless final states can provide information about their velocity
and the ability to discriminate between different $\gd/\X$ masses. More complex final states may be possible as well, but in this proof-of-concept study we restrict ourselves to two body decays only. 

\begin{figure}[t]
	\includegraphics[width = 0.75\linewidth]{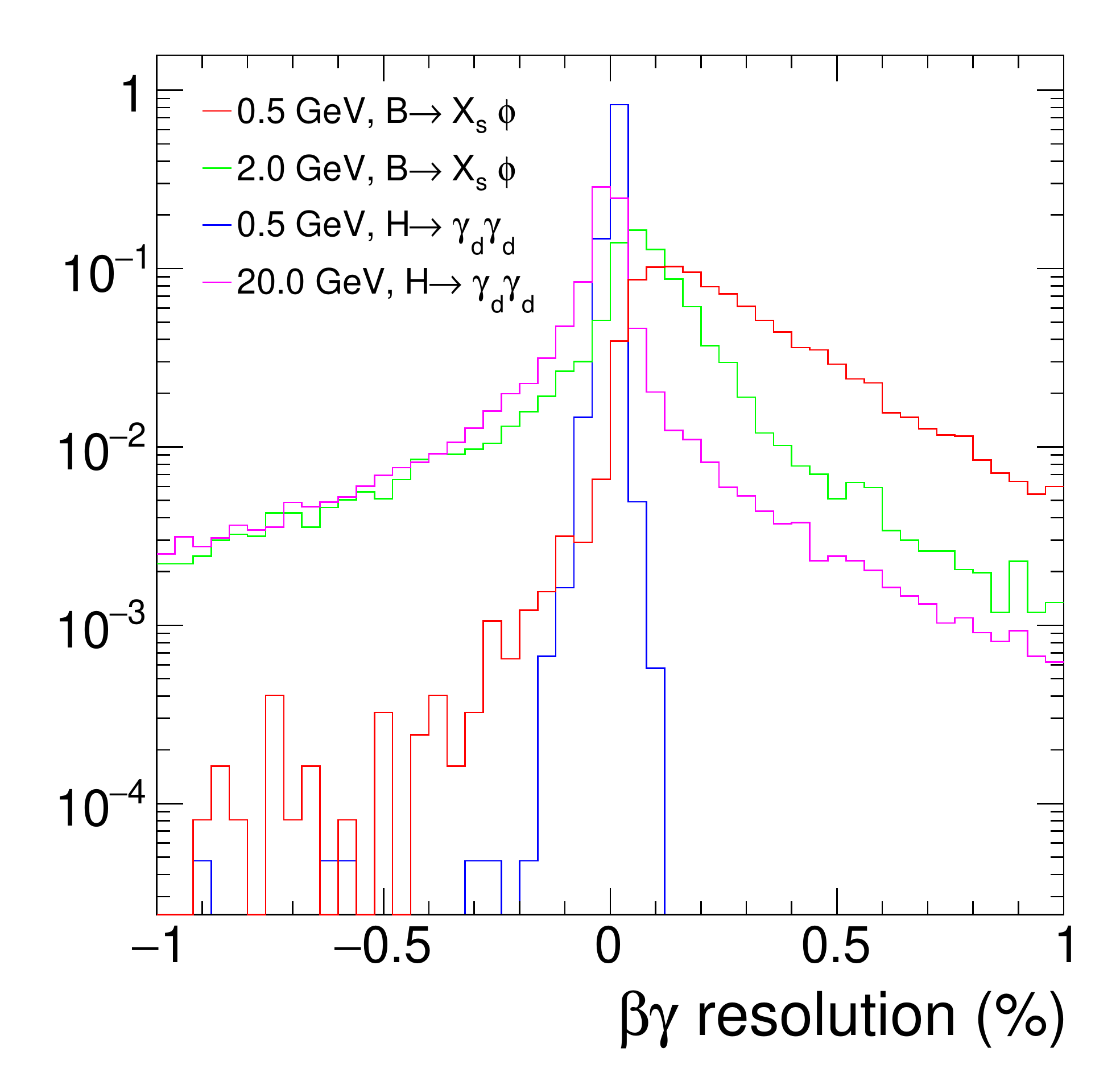}
	\caption{Boost ($\beta\gamma$) resolution for different signals.}
	\label{fig:betares}
\end{figure}

The $\beta$ resolution which can be achieved with our simple tracking layout, using geometric information only and assuming massless final states,
 is shown in Fig.~\ref{fig:betares}. It is Gaussian for the $h \to \gd\gd$ benchmarks, while in the $B\to X_s \X$ case
it is non-Gaussian and biased, because the $\X$ decay products are so slow that the approximation of $\beta=1$
begins to break down for them. Nevertheless we can still reconstruct the $\gd/\X$ velocity to better than 1$\%$ in all cases, 
which in practice means that the ability to discriminate between different $\gd/\X$ masses is largely dominated
by the actual distribution of $\gd/\X$ velocities for a given mass, and not by the detector resolution.

\begin{figure}[t]
	\includegraphics[width = 0.49\linewidth]{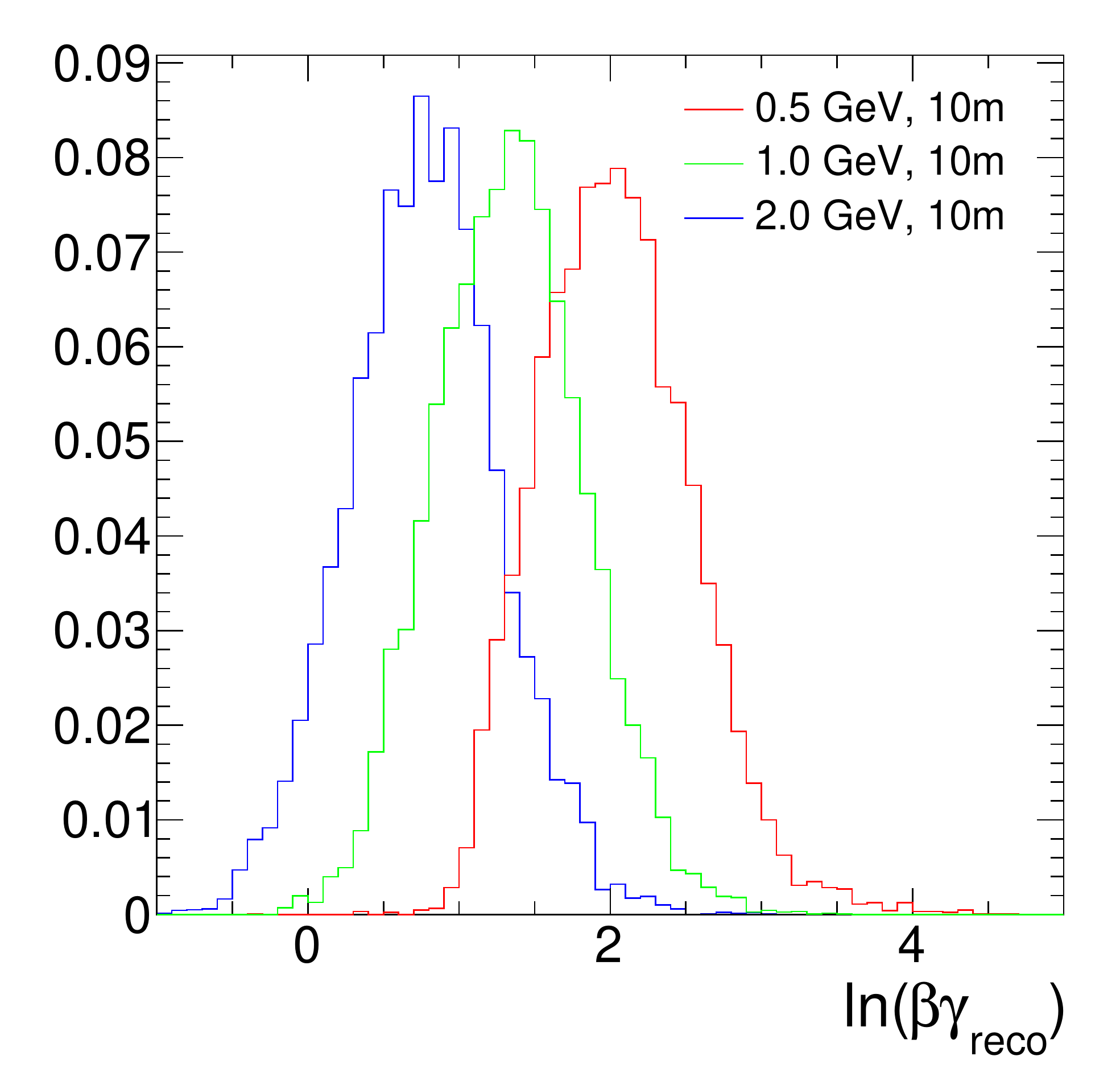}
	\includegraphics[width = 0.49\linewidth]{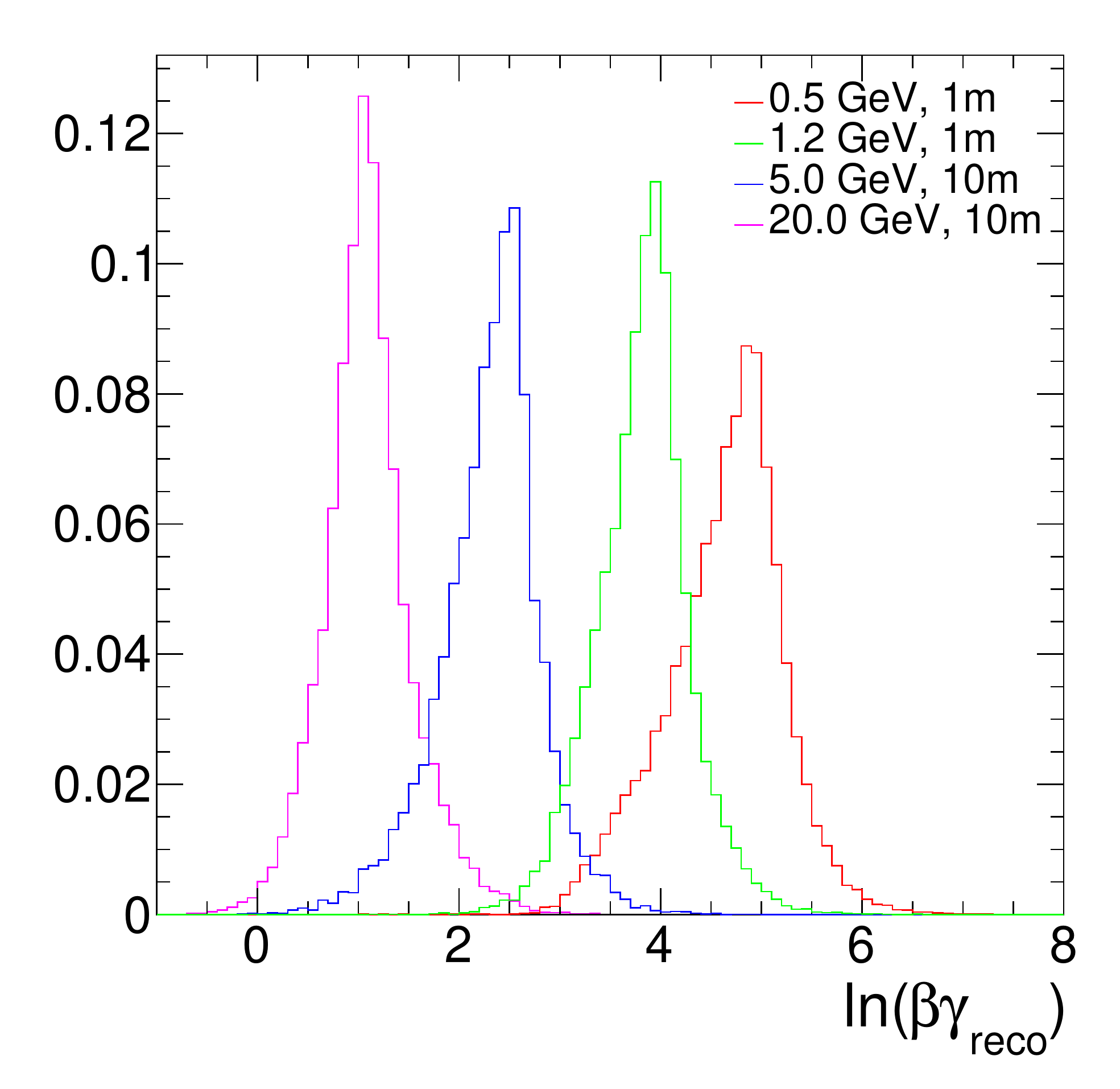}
	\caption{Reconstructed LLP boost for different $B\to X_s\X$ (left) and  $h\to \gd\gd$ (right) mass benchmarks and $c\tau =1$ or $10$\,m. }
	\label{fig:recoboosts}
\end{figure}

The corresponding distribution of reconstructed $\gd/\X$ boosts is shown in Fig.~\ref{fig:recoboosts} for different 
$B\to X_s \X$ and $h\to \gd\gd$ masses and lifetimes. We achieve good discrimination across a wide range of masses
in the $h\to \gd\gd$ case, but perhaps more surprisingly we also have some discriminating power between different
$\X$ masses for the $B\to X_s\X$ benchmark. 

\begin{figure}[t]
	\includegraphics[width = 0.49\linewidth]{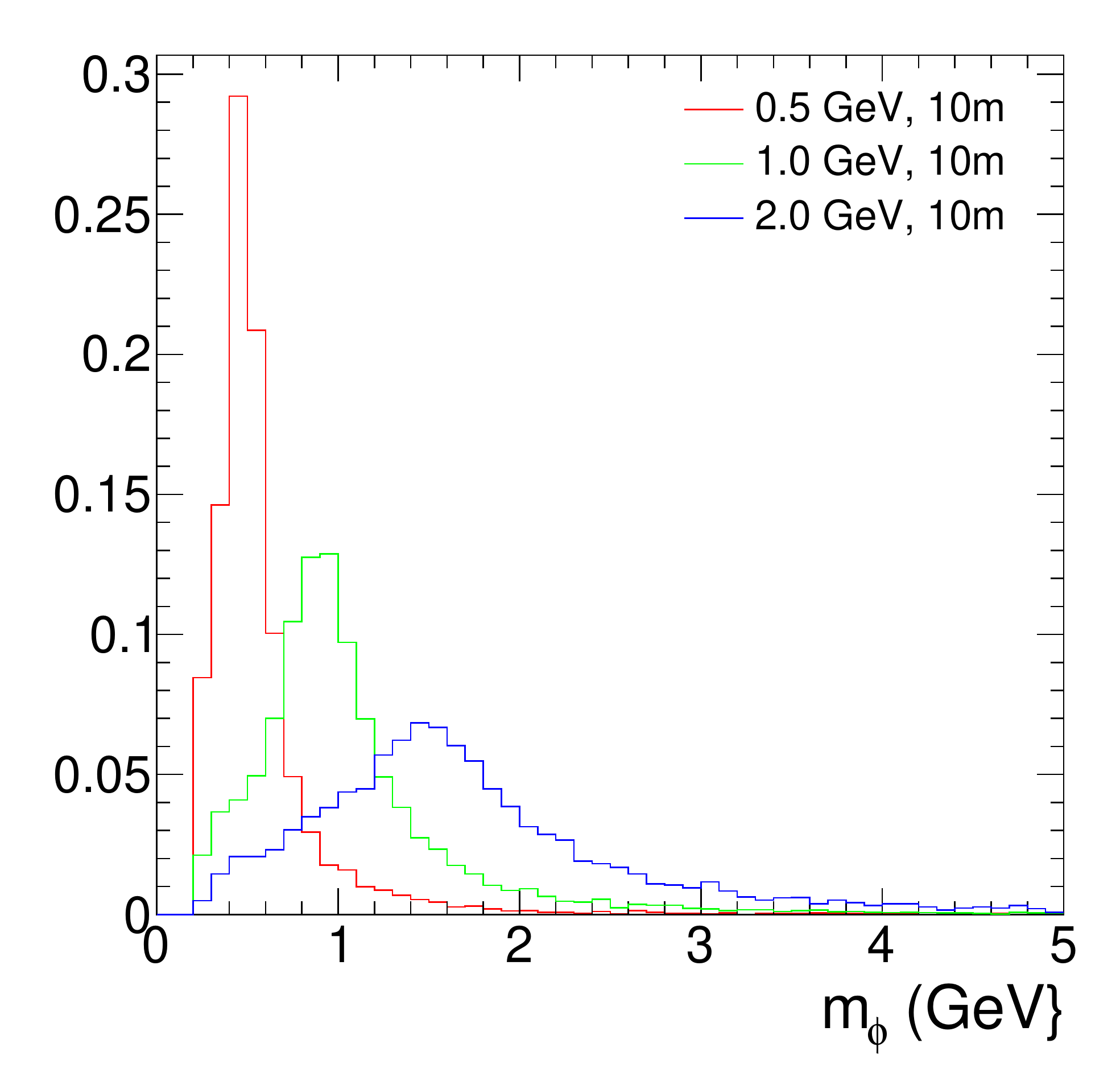}
	\includegraphics[width = 0.49\linewidth]{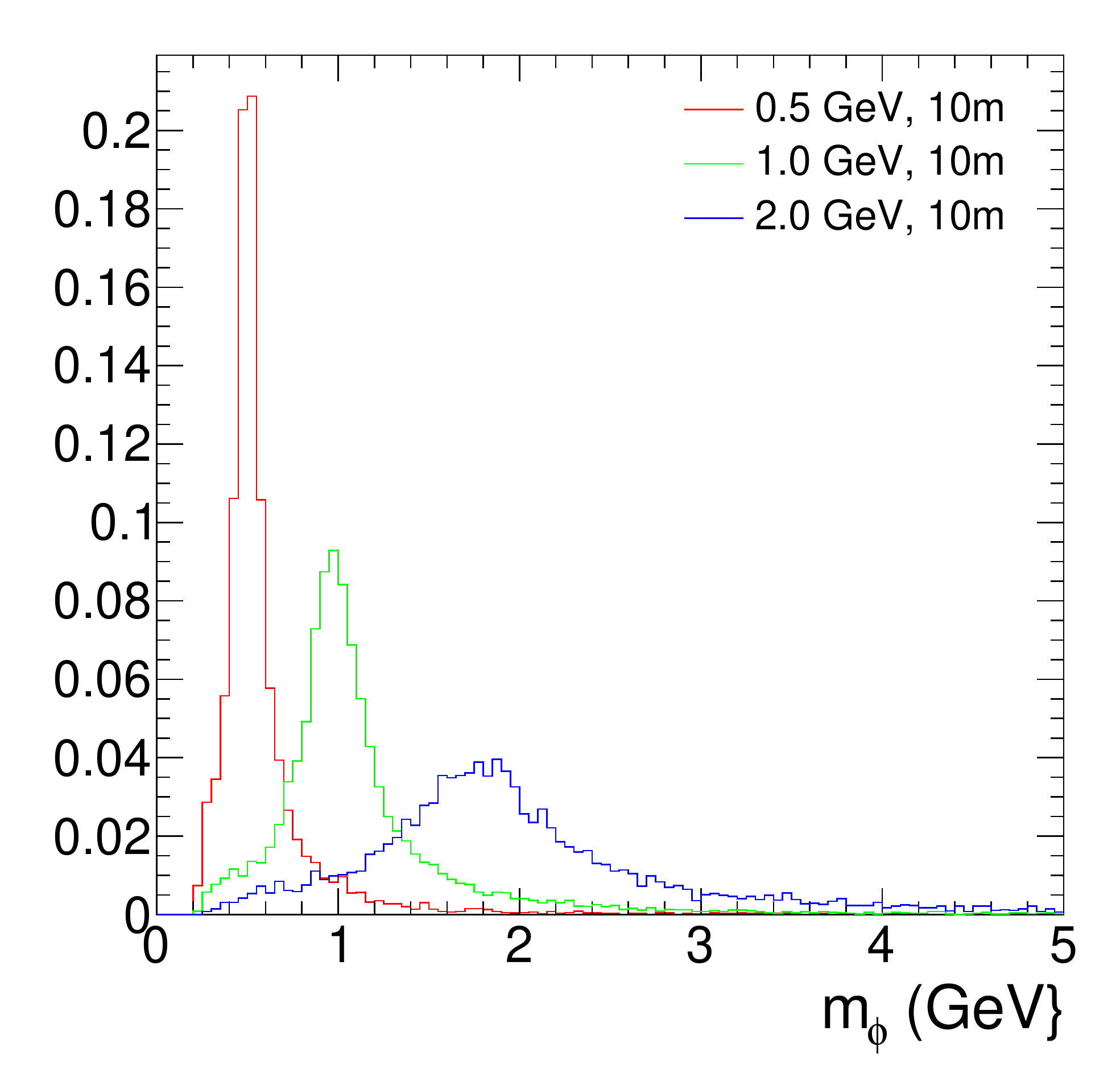}
	\caption{Reconstructed LLP mass for different $B\to X_s\X$ mass benchmarks and $c\tau =10$\,m, for 100~ps (left) and 50~ps (right) time resolution.}
	\label{fig:recomass}
\end{figure}

This approach to a mass measurement uses only spatial information, but complementary information may be provided by using timing 
information from the RPC stations to measure the velocity of the decay products. Such a mass reconstruction can be useful to discriminate between
slow-moving new states, or to veto unexpected slow-moving SM backgrounds which decay or scatter in the box. We show the performance in the $B\to X_s \X$ case in Fig.~\ref{fig:recomass},
assuming timing resolutions of 100~ps or 50~ps for the RPCs, which may be affordably achievable.\footnote{The calibration and time-alignment of the box can be performed with
cosmic rays, just as the initial time alignment of LHCb was.} in the next decade. The 
reconstructed mass resolution depends very strongly on the mass, because heavier $\X$ particles produce decay products which are significantly faster on the scales
relevant here. Nevertheless even in the 100~ps case it is largely possible to separate the 0.5~GeV and 2.0~GeV signals, which also means that we should
be able to reconstruct, e.g., any residual background $K^0_S \to \pi\pi$ decays that might occur in the box.
In the Higgs case the two-body decay products are always too fast to allow this kind of mass reconstruction, though conversely this also allows
them to be discriminated from SM backgrounds and from lighter exotic long-lived states produced, for instance, in heavy flavour decays.

\section{Discussion and conclusions}
\label{sec:llpdetectors}
The results for the $h \to \gd\gd$ and $b \to s \X$ benchmarks provide a proof-of-concept that \expname{} may significantly extend the reach of LLP searches beyond what the current LHC experiments can achieve,
while also covering significant regions of the parameter spaces that other proposed LLP searches seek to explore. However, both for these generic benchmarks and for more specific theories, there is little theoretical
guidance as to where the focus of LLP searches should fall within the wide range of unconstrained LLP masses, branching ratios and lifetimes, and it is difficult to physically construct a single detector capable of exploring 
this entire parameter space.
Thus a picture of complementary detectors with decorrelated backgrounds emerges as the natural choice for a comprehensive LLP search programme.

In this picture, ATLAS and CMS emerge as the most suitable tools to search for heavy and/or charged LLPs for a wide range of lifetimes, due to their $\sim 4\pi$ coverage and rather large size. LHCb adds complementary sensitivity for lower mass LLPs with relatively short lifetimes. Beam dump experiments such as NA62 and SHiP, and ultra-forward LHC detectors like FASER, benefit from reduced backgrounds and a focused LLP source, and can therefore be instrumented with a magnet while also covering a wide range of potential lifetimes. However, 
for LLPs above a few GeV in mass or LLPs which are produced through a heavy portal, they have limited reach, and in particular they can only explore the Higgs sector indirectly. By contrast, collider-based and shielded detectors such as  \expname{} or MATHUSLA can explore further in 
mass, but require larger geometric sizes to ensure adequate fiducial acceptances. This and other practical considerations makes it more difficult to instrument this type of detector with a magnet, so that 
instead a combination of time-of-flight information, possible calorimetry, and kinematic constraints must be used in order to infer the mass and/or boost of any observed signal candidates.
 
The location, size and dominant backgrounds for the latter two proposals are different -- the MATHUSLA detector will nominally achieve both a larger solid 
angle coverage and up to ten times the integrated luminosity compared to \expname{}, but is approximately ten times further from the IP.  Consequently, the absolute reach, the LLP lifetime at their peak
sensitivities, as well as their expected sensitivity to different LLP decay morphologies differs. Further significant differences are expected for the optimal tracker layouts; the potential to deploy more 
complex detector technologies such as calorimetry; alignment strategies; and reconstruction strategies.\footnote{An example of this is that optimization of the tracker layout for \expname{} in 
Sec.~\ref{sec:tracking} requires minimization of the distance between the LLP decay vertex and the first measured point on the trajectories of the decay products. This parameter is particularly 
critical for \expname{} because of the relatively short lever arm between the LLP production and decay vertices.} These differences enhance the complementarity between both experimental setups 
and imply that compatible LLP signals seen in both experiments would represent compelling evidence of physics beyond the Standard Model.

In summary, we have proposed the construction of a new detector element for the LHCb experiment, taking advantage of a soon-to-be available shielded space currently occupied by the data acquisition.  The Compact Detector for Exotics at LHCb (``\expname{}'') has the potential to significantly enhance the new physics reach and capabilities of the LHCb program, in ways complementary to, and in some cases exceeding, the reach of ATLAS and CMS and the LHCb main detector. We have verified in this proof-of-concept study that efficient vertex reconstruction can be achieved with an RPC tracking implementation, and that reconstruction of LLP mass and boost is feasible. In terms of reach, we project that for Higgs decays to sub-GeV dark photons and for exotic $B$-decays through a Higgs portal, \expname{} would extend the reach of the currently planned LHC program by over one and two orders of magnitude, respectively.

\begin{acknowledgments}
We thank Yasmine Amhis, Bob Cahn, Xabier Cid Vidal, David Curtin, Jared Evans, Tim Gershon, Heather Gray, Gaia Lanfranchi, Jacques Lefrancois, Alex Marks-Bluth, Simone Pagan-Griso, Chris Parkes, Giovanni Passaleva, Michael Sokoloff, Frederic Teubert and Mike Williams for helpful discussions. We especially thank Tim Gershon and Mike Williams for comments on the manuscript and Rolf Lindner for useful discussions about the layout of the LHCb cavern and for providing us with high resolution drawings of it.  VVG acknowledges funding from the European Research Council (ERC) under the European Union's Horizon 2020 research and innovation programme under grant agreement No 724777 ``RECEPT''. SK and MP are supported in part by the LDRD program of LBNL under contract DE-AC02-05CH11231, and by the National Science Foundation (NSF) under grants No. PHY-1002399 and PHY-1316783. DR acknowledges support from the University of Cincinnati. SK, MP and DR also thank the Aspen Center of Physics, supported by the NSF grant 
PHY-1066293, where this paper was completed.  This research used resources of the National Energy Research Scientific Computing Center, which is supported by the Office of Science of the DoE under Contract No.~DE-AC02-05CH11231.
\end{acknowledgments}

\bibliographystyle{apsrev4-1}

\end{document}